# Observation of Superconductivity Induced Ferromagnetism in an Fe-Chalcogenide Superconductor

Nathan J. McLaughlin[1], Hailong Wang[2], Mengqi Huang[1], Eric Lee-Wong[1,3], Lunhui Hu[1], Hanyi Lu[1], Gerald Q. Yan[1], G. D. Gu[4], Congjun Wu[1,5,6], Yi-Zhuang You[1], Chunhui Rita Du[1,2,*]

[1]Department of Physics, University of California, San Diego, La Jolla, California 92093, USA
[2]Center for Memory and Recording Research, University of California, San Diego, La Jolla, California 92093, USA
[3]Department of NanoEngineering, University of California, San Diego, La Jolla, California 92093, USA
[4]Condensed Matter Physics and Materials Science Division, Brookhaven National Laboratory, Upton, New York 11973, USA
[5]School of Science, Westlake University, Hangzhou, Zhejiang 310024, China
[6]Institute of Natural Sciences, Westlake Institute for Advanced Study, Hangzhou, Zhejiang 310024, China

*Correspondence to: c1du@physics.ucsd.edu

**Abstract**: The interplay among topology, superconductivity, and magnetism promises to bring a plethora of exotic and unintuitive behaviors in emergent quantum materials. The family of Fe-chalcogenide superconductors $FeTe_xSe_{1-x}$ are directly relevant in this context due to their intrinsic topological band structure, high-temperature superconductivity, and unconventional pairing symmetry. Despite enormous promise and expectation, the local magnetic properties of $FeTe_xSe_{1-x}$ remain largely unexplored, which prevents a comprehensive understanding of their underlying material properties. Exploiting nitrogen vacancy (NV) centers in diamond, here we report nanoscale quantum sensing and imaging of magnetic flux generated by exfoliated $FeTe_xSe_{1-x}$ flakes, providing clear evidence of superconductivity-induced ferromagnetism in $FeTe_xSe_{1-x}$. The coexistence of superconductivity and ferromagnetism in an established topological superconductor opens up new opportunities for exploring exotic spin and charge transport phenomena in quantum materials. The demonstrated coupling between NV centers and $FeTe_xSe_{1-x}$ may also find applications in developing hybrid architectures for next-generation, solid-state-based quantum information technologies.

# INTRODUCTION

Topological superconductors harboring exotic Majorana fermions have received intensive research interest over the past several years due to their significant potential for developing transformative, fault-tolerant quantum-computing paradigms.[1,2] This emergent class of superconductors is distinguishable from its conventional counterparts by its topologically protected surface or edge state, as well as exhibiting a range of exotic and highly interesting quantum phenomena.[1,3] Among the potential material candidates, the $FeTe_xSe_{1-x}$ family naturally stands out due to its high superconducting transition temperature, simple crystal structure, and intrinsic topological band structure due to the bulk band-inversion induced by spin-orbit coupling.[4–7]

Over the past few years, pioneering research efforts have gone towards experimentally investigating the topological nature and the superconducting properties of $FeTe_xSe_{1-x}$.[4,6,8] However, the topological superconductivity observed in these experiments is a consequence of the proximity effect from the bulk superconductivity to the surface Dirac state. It is natural to examine whether the surface superconductivity can provide important information about the bulk gap function symmetry. In particular, time-reversal symmetry as well as parity and charge conjugation are three fundamental discrete symmetries of interest. The existence or lack of time-reversal symmetry breaking (TRSB) is an important question for exploring pairing mechanisms in strongly correlated superconductors. TRSB pairing symmetries have been theoretically proposed in iron-based superconductors.[9] Experimentally, thermal transport and angle-resolved photoemission spectroscopy studies suggest the spontaneous TRSB and opening of a band gap at the surface Dirac cone in superconducting $FeTe_xSe_{1-x}$,[10,11] which indicates the presence of ferromagnetism induced by an intrinsic unconventional pairing symmetry.[12] The interplay among topology,[4,6,8,13] superconductivity,[4,7] and magnetism[10,11] makes $FeTe_xSe_{1-x}$ an attractive platform to explore a range of emergent quantum spin and charge transport phenomena.[11,14–16] Despite remarkable progress, detailed knowledge of the local magnetic properties of superconducting $FeTe_xSe_{1-x}$ remains elusive, and direct evidence to corroborate the coexistence of superconductivity and ferromagnetism in $FeTe_xSe_{1-x}$ is still lacking. The major difficulty results from the fact that the magnetic flux generated by superconductors is often shielded by the Meissner effect,[17] which is challenging to access using conventional magnetometry methods.

Here, we introduce nitrogen-vacancy (NV) centers,[18,19] optically active atomic defects in diamond that act as single-spin sensors, to perform nanoscale quantum imaging and sensing of magnetic flux generated by an exfoliated $FeTe_{0.7}Se_{0.3}$ (FTS) flake.[11] By performing NV Rabi oscillation measurements,[20,21] we directly image the spatial distribution of supercurrents in FTS. By measuring magnetic stray fields and fluctuations by the NV optical detection of magnetic resonance (ODMR) and relaxometry techniques,[22–26] we provide clear evidence of superconductivity induced ferromagnetism in FTS.[12] The presented NV quantum sensing and imaging platform operates in an accessible, table-top format, which can be extended naturally to a large family of two-dimensional van der Waals materials,[27] providing new opportunities for investigating local electrical and magnetic behaviors in emergent quantum materials.

# EXPERIMENTAL DETAILS

Figure 1a shows a schematic of our measurement platform, where a 140-nm-thick FTS flake with lateral dimensions of ~11 μm × ~9 μm is transferred onto a single crystal diamond substrate containing NV centers implanted ~5 nm below the surface. The density of NV centers is characterized to be ~1500/μm$^2$, providing a convenient platform to achieve wide-field imaging based on the ensemble of NV spins.[28–30] An on-chip Au stripline is fabricated on the diamond sample, allowing the application of microwave currents to control the quantum spin state of the NV centers. A photoluminescence image [Fig. 1b] shows an overview of a prepared NV-FTS device, where an FTS flake is located ~4 μm from the Au stripline. The exfoliated FTS flake exhibits the characteristic superconducting phase transition at $T_c$ =14.5 K, as shown in Fig. 1c, in agreement with the previous electrical transport study (see Supporting Information Section 1 for details).[10]

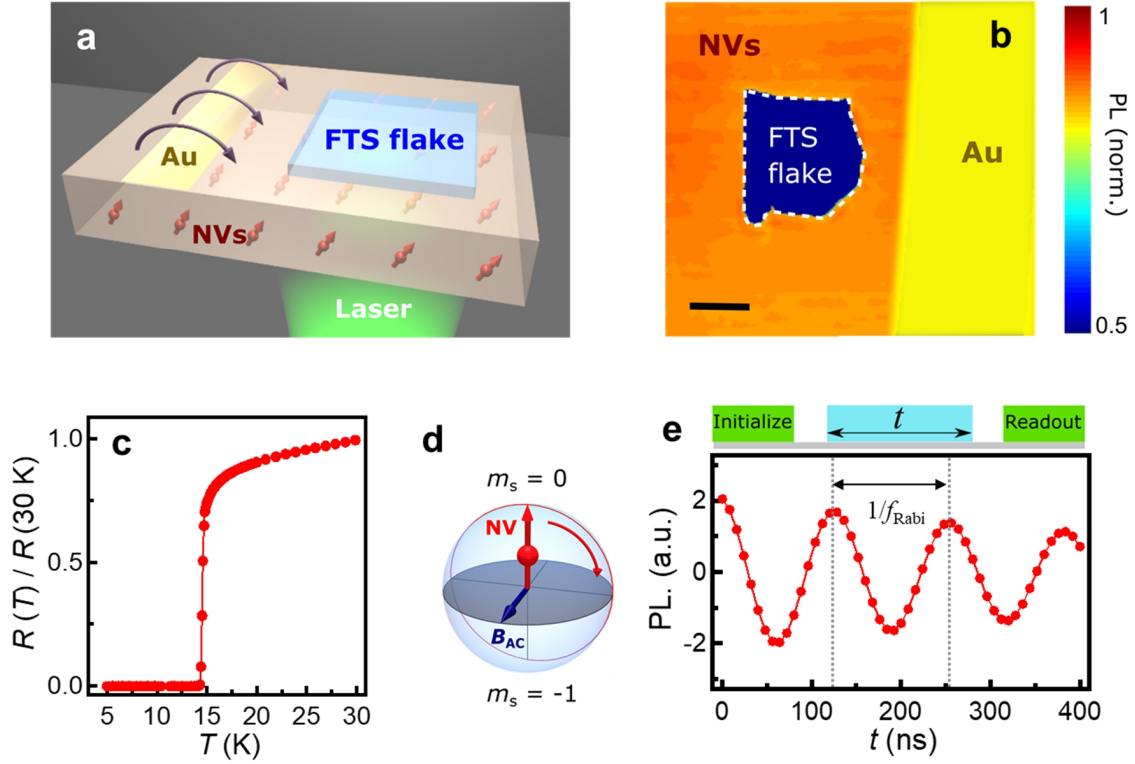

**Figure 1.** Measurement of magnetic flux generated by an exfoliated FTS flake by NV centers. (a) Schematic of an exfoliated FTS flake transferred onto a diamond membrane for NV wide-field magnetic imaging measurements. (b) Photoluminescence image showing an overview of a prepared NV-FTS device. The scale bar is 5 μm. (c) Electrical transport measurements show a characteristic superconductive phase transition at 14.5 K for an exfoliated FTS flake. (d), (e) Schematic of Rabi oscillations on the Bloch sphere and an experimental measurement spectrum of NV Rabi oscillation. The top panel of (e) shows the optical and microwave sequence of NV Rabi oscillation measurements.

We first employed NV wide-field microscopy[28–30] to perform Rabi oscillation measurements to image circular AC supercurrents in the FTS flake. An NV center is formed by a nitrogen atom adjacent to a carbon atom vacancy in one of the nearest neighboring sites of a diamond crystal lattice.[19] The negatively charged NV state has an $S = 1$ electron spin and serves as a three-level quantum system. When a microwave magnetic field at the electron spin resonance (ESR) frequency is applied to the NV site, the NV spin will oscillate periodically between two different states, e. g. those with m$_s$ = 0 and m$_s$ = −1 in the Bloch sphere, as illustrated in Fig. 1d.

These are referred to as Rabi oscillations.[21,31] Because the $m_s = \pm 1$ spin states of NV centers are more likely to be trapped by a non-radiative pathway back to the $m_s = 0$ ground state and emit reduced photoluminescence,[19] Rabi oscillations of the NV spin can be optically accessed by measuring the spin-dependent photoluminescence. The top panel of Fig. 1e shows the optical and microwave sequence for Rabi oscillation measurements by an NV wide-field microscopy. A 1 µs long green laser pulse is first applied to initialize the NV spin to the $m_s = 0$ state. A microwave pulse at the NV ESR frequency is applied to induce the NV spin transition followed by a second green laser pulse to measure the spin-dependent photoluminescence of NV centers and to re-initialize the NV spins for the next measurement sequence. The microwave pulse is applied 700 ns after turning off the initialization green laser pulse to minimize the laser-induced heating and electric/magnetic excitations. The entire measurement sequence is repeated by 15,000 times during one camera exposure period and the duration of the microwave pulse systematically varies from zero to a few hundred nanoseconds at different camera exposure periods in order to detect a time-dependent variation of NV photoluminescence. The bottom panel of Fig. 1e shows a characteristic

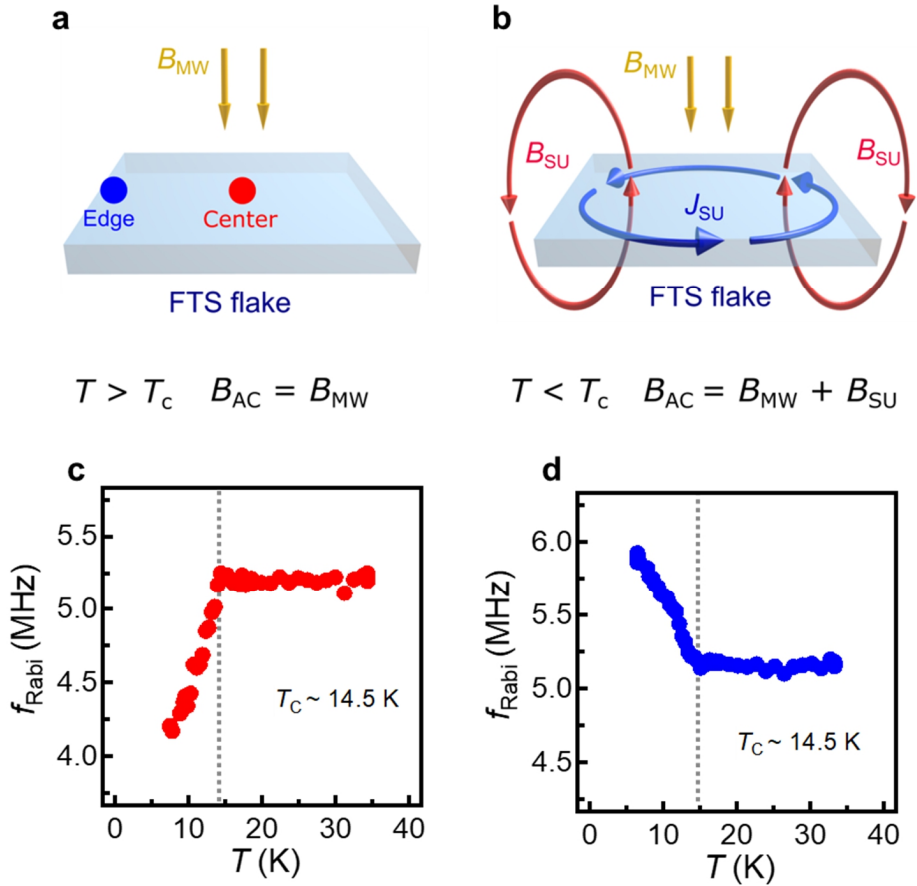

**Figure 2.** Local quantum sensing of superconducting phase transition of an FTS flake. (a), (b) Schematic of the spatial profile of the microwave magnetic field environment of an FTS flake above and below $T_c$. (c), (d) Temperature dependence of Rabi oscillation frequency $f_{Rabi}$ of NV centers located underneath the central (c) and edge (d) areas of the FTS flake. The Meissner screening effect reduces (increases) the effective microwave field in the central (edge) area of the FTS flake, leading to the opposite variations of $f_{Rabi}$ across $T_c$.

NV Rabi oscillation spectrum. The amplitude of the applied microwave magnetic field $B_{AC}$ can be obtained by the frequency of the Rabi oscillation $f_{Rabi}$ as follows: $B_{AC} = \frac{\sqrt{3} f_{Rabi}}{\gamma}$, where $\gamma$ is the gyromagnetic ratio of an NV spin (see Supporting Information Section 2 for details).

In the prepared NV-FTS device, the microwave magnetic field $B_{AC}$ at individual NV sites contains two potential contributions: the oscillating Oersted field $B_{MW}$ produced by the microwave current flowing in the on-chip Au stripline and the magnetic field $B_{SU}$ generated by supercurrents in the FTS flake, as illustrated in Figs. 2a and 2b. When $T > T_c$, the Rabi oscillation of NV centers is only driven by the Oersted field $B_{MW}$. When temperature falls below $T_c$, circular supercurrents in FTS modify the spatial distribution of the microwave magnetic field surrounding the FTS flake.

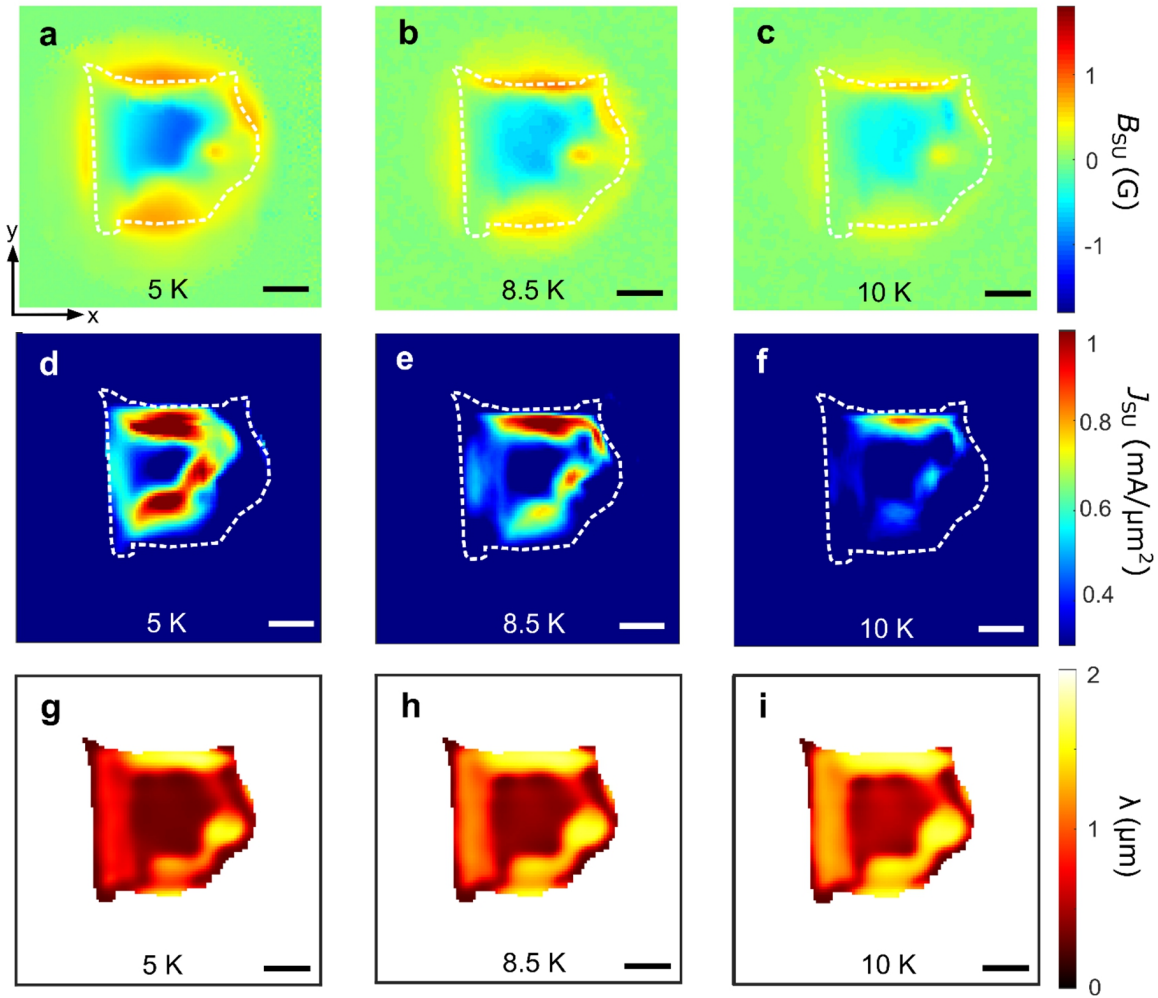

**Figure 3.** NV wide-field imaging of the internal microwave magnetic field, supercurrents, and London penetration depth of an FTS flake. (a)-(c) 2D imaging of the internal microwave magnetic field $B_{SU}$ generated by supercurrents in the FTS flake at 5 K, 8.5 K, and 12 K. (d)-(f) Reconstructed spatial distribution of supercurrents in the FTS flake at 5 K, 8.5 K, and 12 K. (g)-(i) 2D maps of London penetration depth $\lambda$ reconstructed by machining learning analysis with an assumption of a **spatially dependent microwave conductivity**. For all panels, the white dashed line marks the boundary of the exfoliated FTS flake and scale bar is 3 μm.

In the central area of the FTS flake, the microwave field $B_{SU}$ generated by supercurrents is in the opposite direction of the external microwave field $B_{MW}$, effectively reducing the NV Rabi oscillation rate $f_{Rabi}$. $B_{SU}$ and $B_{MW}$ follow the same direction in the boundary area of the FTS flake, leading to an enhancement of $f_{Rabi}$. Figures 2c and 2d show the temperature dependence of the measured Rabi oscillation frequency $f_{Rabi}$ for NV centers underneath the central and boundary areas of the FTS flake, respectively. Notably, $f_{Rabi}$ exhibits a sudden decrease (increase) when temperature falls below the $T_c$ of FTS, in agreement with the picture of the Meissner screening effect discussed above.[31,32] By subtracting the external microwave field generated by the on-chip Au stripline, the internal magnetic field $B_{SU}$ due exclusively to the circulating supercurrents in the FTS flake can be extracted (see Supporting Information Section 2 for details).

Figures 3a-3c show 2D maps of $B_{SU}$ measured at 5 K, 8.5 K, and 10 K. When $T$ = 5 K, magnetic flux expulsion due to the Meissner effect[32,33] leads to a negative magnetic field in the central area of the flake and a positive magnetic field in the edge area. Note that the sign of $B_{SU}$ represents its orientation relative to the external microwave field. The magnitude of $B_{SU}$ decreases with increasing temperature and eventually vanishes above $T_c$. The spatially resolved $B_{SU}$ allows us to reconstruct the distribution of supercurrent $J_{SU}$ in the FTS flake, as shown in Figs. 3d-3f (see Supporting Information Section 3 for details).[28,29] The supercurrent forms a circular loop with a maximum density of ~1 mA/μm$^2$. It gradually decays and approaches zero in the central region of the FTS flake. The length scale that the supercurrents decay inside a superconductor is determined by the London penetration depth $\lambda$.[31] Microscopically, $\lambda$ is associated with the imaginary part of the electrical conductivity and the superfluid density of FTS.[31] Using a machine learning analysis model,[34] we fit the spatial distribution of the microwave field $B_{SU}$ to obtain 2D maps of London penetration depth, as shown in Figs. 3g-3i (see Supporting Information Section 4 for details). The spatial variation of $\lambda$ within the exfoliated FTS flake demonstrates the existence of superconducting domains possibly induced by inhomogeneities or defects.[6]

In conventional *s*-wave superconductors, ferromagnetism is usually prohibited because the characteristic spin-singlet *s*-wave pairing rule dictates antiparallel spin configuration of cooper pairs. Thus, the emergence of superconductivity-induced ferromagnetism is usually accompanied with the spin triplet states or unconventional superconducting pairing symmetries such as experimentally observed in UPt$_3$,[35] UTe$_2$,[36] Ba$_{1-x}$K$_x$Fe$_2$As$_2$,[37] and recently theoretically predicted in FTS.[12] Next, we employ the NV ODMR technique[38–40] to detect magnetic stray fields generated by a superconducting FTS flake, providing direct evidence to the superconductivity-induced ferromagnetism. Figure 4a shows a schematic of our measurement geometry. An external magnetic field $B_{ext}$ ~ 60 G is applied with an angle of 39 degrees relative to the normal of the sample plane. The in-plane projection of $B_{ext}$ is approximately along the *x*-axis direction. Note that both supercurrents and magnetization in the FTS flake can potentially produce a sizeable stray field. If the superconductivity-induced ferromagnetism is spontaneously aligned to the direction of the external magnetic field, the out-of-plane component of the stray field generated by a tilted ferromagnetization is negligibly small in the central area of the FTS flake and exhibits opposite signs at the + *x* and − *x* boundaries, as illustrated in Fig. 4a. In contrast, the out-of-plane component of the internal field generated by supercurrents follows the characteristic Meissner effect, exhibiting a decrease in the center and an increase on the boundary of the FTS flake, as shown in Fig. 2b. Based on these two distinct spatial geometries, we can qualitatively separate these two effects.

The top panel of Fig. 4b shows the microwave and optical sequence of pulsed NV ODMR measurements. We utilize 1 μs long green laser pulses for NV initialization and readout and 100 ns long microwave π pulses to induce NV spin transitions. The entire sequence is repeated ~15,000 times during one camera exposure period and the frequency $f$ of the microwave pulses is swept for different camera exposure periods. When $f$ matches the NV ESR conditions, reduced photoluminescence is observed. Note that the NV ensemble used in our study has four pairs of ESR transitions in one ODMR spectrum in response to an arbitrary magnetic field, allowing extraction of the out-of-plane component of the internal stray field $B_z$ generated by the FTS flake (see Supporting Information Section 5 for details). Figures 4b and 4f show the 2D maps of $B_z$ measured at 5 K and 12 K, respectively. When $T < T_c$, $B_z = B_{SU} + B_{FM}$, where $B_{SU}$ and $B_{FM}$ represent the out-of-plane fields produced by supercurrents and ferromagnetism in the FTS flake, respectively. At 5 K, $B_z$ is dominated by the Meissner screening effect, showing a positive value at the boundary area of the FTS flake and a negative value in the central region. While the contribution of $B_{FM}$ is heavily masked in this situation, the asymmetry of $B_z$ along the $x$-axis direction indicates the existence of a tilted magnetization in the FTS flake. When $T = 12$ K in proximity to $T_c$, the Meissner effect is significantly suppressed due to the reduced supercurrents, thus, the asymmetry of $B_z$ becomes more pronounced. The sign of $B_z$ reverses at the two opposite boundaries of the FTS flake, providing clear evidence to the induced magnetic moment in superconducting FTS.

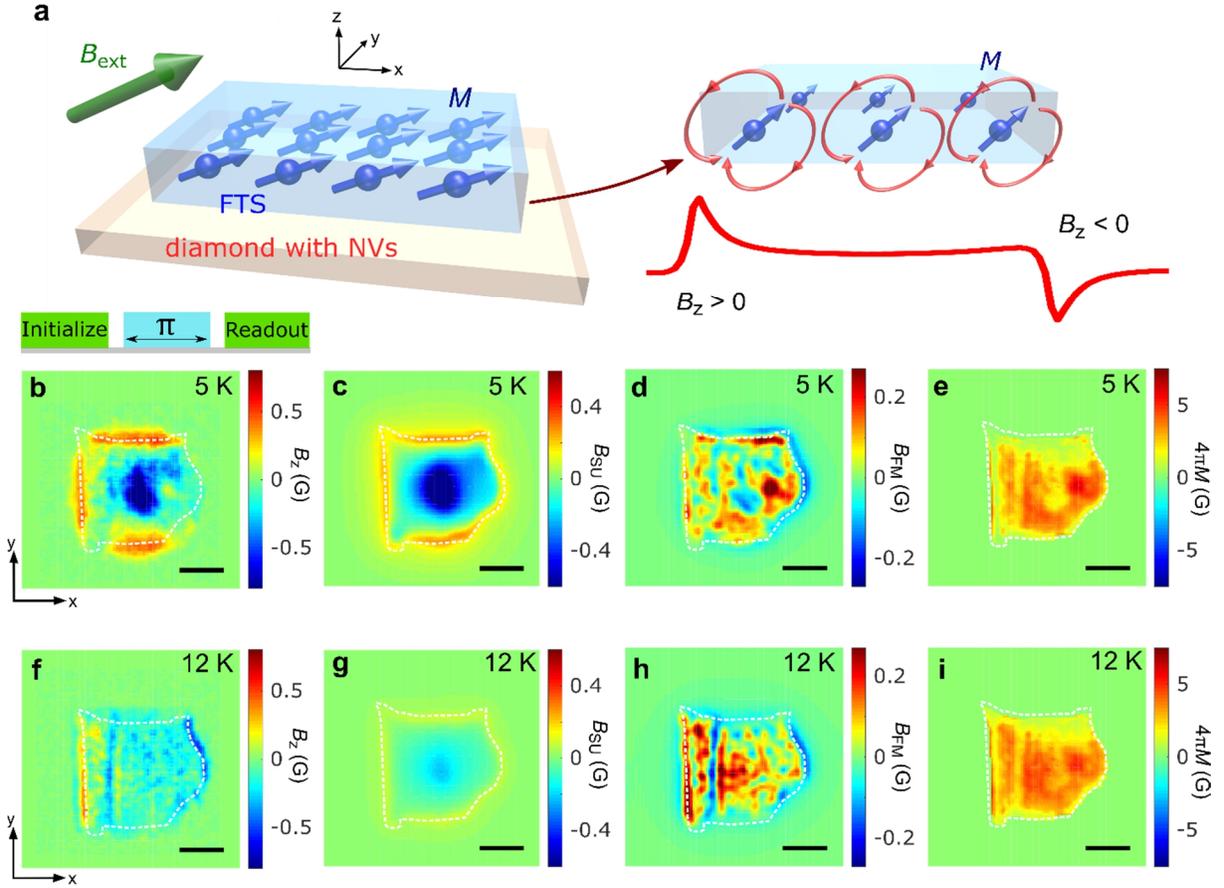

**Figure 4.** NV wide-field imaging of superconductivity-induced ferromagnetism in FTS. (a) Experimental measurement geometry to detect superconductivity-induced ferromagnetism in an FTS flake. An external magnetic field $B_{ext}$ is applied to tilt the magnetization away from the equilibrium position (out-of-plane direction, $z$-axis). The in-plane projection of $B_{ext}$ is approximately along the $x$-axis direction. The out-of-plane component of the internal DC magnetic field generated by the ferromagnetism shows the sign reversal feature at the $+x$ and $-x$ boundaries of the FTS flake. (b), (f) 2D wide-field imaging of internal magnetic field $B_z$ generated by the FTS flake at 5 K (b) and 12 K (f). Top panel of (b): optical and microwave sequence of NV ODMR measurements. (c), (g) Extracted 2D images of stray field $B_{SU}$ produced by supercurrents at 5 K (c) and 12 K (g). (d), (h) Extracted 2D images of stray field $B_{FM}$ generated by the superconductivity-induced ferromagnetism at 5 K (d) and 12 K (h). (e), (i) Reconstructed spatially dependent magnetization of the superconducting FTS flake at 5 K (e) and 12 K (i). For figures from (b) to (i), the white dashed line marks the boundary of the exfoliated FTS flake and scale bar is 3 μm.

To quantitatively separate the two mutually entangled effects, we performed machine-learning assisted simulations to reconstruct the spatially dependent magnetic field $B_{SU}$ and $B_{FM}$ (see Supporting Information Section 6 for details). Figures 4c, 4d, 4g, and 4h show the extracted 2D maps of $B_{SU}$ and $B_{FM}$ at 5 K and 12 K, respectively. The contribution from the supercurrents (Figs. 4c and 4g) largely resembles the Meissner effect and gets significantly reduced when temperature approaches $T_c$. Notably, the obtained 2D map of $B_{FM}$ follows the "sign reversal" feature at the $+x$ and $-x$ boundaries of the FTS flake, suggesting that the induced magnetic moment follows the direction of the external magnetic field. It is worth mentioning that $B_{FM}$ exhibits a nonuniform distribution, indicating a multidomain structure [Figs. 4d and 4h] of the superconductivity-induced ferromagnetism in the FTS flake, which can be attributable to defects, inhomogeneity and domain wall pinning effects.[40] When the in-plane component of the external magnetic field $B_{ext}$ is rotated by ~130 degrees, the observed $B_{SU}$ largely remains the same while the field pattern of $B_{FM}$ rotates accordingly (see Supporting Information Section 6 for details), confirming the weak anisotropy of the superconductivity-induced ferromagnetism whose orientation is tunable by the external magnetic field. Based on the spatial-dependent $B_{FM}$, we further reconstructed the 2D profile of the magnetization $4\pi M$ of FTS, as shown in Figs. 4e and 4i (see Supporting Information Section 6 for details). The obtained $4\pi M$ exhibits spatially dependent variation with an average value of ~4 G for $T < T_c$. Theoretically, the upper limit of $M$ of superconducting FTS can be estimated by assuming that all the charge carriers are polarized: $M_{max} = \frac{b\mu_B}{V}$, where $b\mu_B$ is the magnetization of each Fe atom, $\mu_B$ is the Bohr magneton, and $V$ is the average volume containing one Fe cation. With $b = 0.0464$ and $V = 1.0 \times 10^{-28}$ m$^3$,[12] $4\pi M_{max}$ is calculated to be 15 Oe, in qualitative agreement with our experimental results. We highlight that the measured spin susceptibility of FTS is ~0.1 at 12 K, which is about 200 times larger than the value measured in the paramagnetic state (see Supporting Information Section 1 for details), suggesting that the observed magnetic moment cannot be a result of spin polarization induced by the external magnetic field. In addition, we also performed NV measurements of the stray field $B_z$ produced by FTS above $T_c$. The measured magnetic field falls below our detection limit over entire measurement region at 16 K (see Supporting Information Section 6 for details). All these results demonstrate that the observed ferromagnetism is indeed driven by the superconductivity in FTS. In addition to DC stray fields, the superconductivity induced ferromagnetism in FTS also generates fluctuating magnetic fields at the characteristic frequencies of thermal magnons. When the frequency of FTS magnons matches the NV ESR frequency, the magnetic fluctuations will induce spin transitions of a proximal NV center. Our experimentally measured NV relaxometry results are well correlated to the field-dependent variation of the magnon density of FTS, providing another piece of evidence to the superconductivity-induced ferromagnetism in FTS (see Supporting Information Section 7 for details).[22,41]

Next, we briefly discuss the physical origin of the observed superconductivity-induced ferromagnetism. In FTS, the superconducting carriers are electrons and holes near the Fermi level, and localized magnetic moments can be carried by the $d$ orbitals of Fe atoms,[42] impurities, or defects. Due to the weak exchange coupling of carriers at the Fe sites, long-range ferromagnetic ordering is absent in the normal state. When $T < T_c$, TRSB accompanied by unconventional pairing symmetries could take place, leading to the emergence of ferromagnetism in FTS. Owing to the multi-orbital nature of the Fe atoms,[12] there may exist nearly degenerate gap function symmetries. To reduce the free energy, they often prefer a superposition with a relative $\pm\frac{\pi}{2}$ phase difference in the mean-field theory. The ferromagnetic order is inversion symmetric, hence, it can only be induced by mixing two pairing orders with the same parity. The mixing between $A_{1g}$ and $A_{2g}$, or $B_{1g}$ and $B_{2g}$ can yield the ferromagnetic order along the $z$-direction. On the other hand, the in-plane component of ferromagnetism could be a consequence of mixing a pairing symmetry of $A$ or $B$-type, with a two-dimensional representation of the $E$-type.[12] Loosely speaking, $A$, $B$ and $E$ refer to the $s$, $d$ and $p$-wave pairing symmetries respectively; the subindices 1 and 2 refer to the parity with respect to the vertical plane reflection; and $g$ means even parity under inversion. Due to the orbital degree of freedom, the $E$-type pairing symmetry can also be spin-singlet. Such a complex pairing gap function can couple to the bulk magnetization via spin-orbit coupling according to a symmetry analysis,[12] enabling the establishment of a long-range magnetic order in superconducting FTS. This driving force is due to the TRSB Cooper pairing, whose effect may be weak. Nevertheless, the FTS system contains local moments,[42] and they may be further aligned by the bulk magnetization to amplify such effect.

In summary, we have demonstrated NV-based quantum sensing and imaging of magnetic flux generated by exfoliated FTS flakes. The obtained spatial profile of stray fields generated by supercurrents in FTS flakes follows the characteristic Meissner screening effect. By employing NV ODMR method, we provide direct evidence of superconductivity-induced ferromagnetism in FTS, opening up new opportunities for exploring the rich physics related to the interplay among topology, superconductivity, and magnetism in topological superconductors. At the same time, our results also reveal the unconventional pairing mechanism and the time reversal symmetry breaking in FTS, demonstrating NV centers as a versatile local probe to investigate nanoscale electrical and magnetic properties of emergent quantum materials. The observed spatially tunable coupling between NV centers and superconducting FTS also provides a new route for developing hybrid quantum architectures,[43] aiding in the development of next-generation, solid-state-based quantum information technologies.

**Acknowledgements**. Authors would like to thank Eric E. Fullerton for providing the Physical Property Measurement System for the electrical transport characterizations and Yuxuan Xiao for assistance in sample preparation. Authors thank Peter Johnson, John Tranquada, Ruolan Xue, Amir Yacoby, and Mark Ku for insightful discussions. N. J. M., H. L. and C. H. R. D. were supported by the Air Force Office of Scientific Research under award FA9550-20-1-0319 and its Young Investigator Program under award FA9550-21-1-0125. E. L.-W., G. Q. Y. and C. H. R. D. acknowledge the support from U. S. National Science Foundation under award ECCS-2029558 and DMR-2046227. The work at BNL was supported by the US Department of Energy, office of Basic Energy Sciences, contract no. DOE-sc0012704.